\documentclass[reprint,amsmath,amssymb,aip,jcp,superscriptaddress]{revtex4-1}

\usepackage{graphicx}
\usepackage{dcolumn}
\usepackage{mathtools}
\usepackage{color}

    \newcommand{\neqref}[1]{(\ref{#1})}

\begin{document}

\title{Analytic solution of the SEIR epidemic model via asymptotic approximant}
\author{Steven J. Weinstein}
\affiliation{School of Mathematical Sciences, Rochester Institute of Technology, Rochester, NY 14623, USA}
\affiliation{Department of Chemical Engineering, Rochester Institute of Technology, Rochester, NY 14623, USA}
\author{Morgan S. Holland}
\affiliation{School of Mathematical Sciences, Rochester Institute of Technology, Rochester, NY 14623, USA}
\author{Kelly E. Rogers}
\affiliation{School of Mathematical Sciences, Rochester Institute of Technology, Rochester, NY 14623, USA}
\author{Nathaniel S. Barlow}
\email{nsbsma@rit.edu}
\affiliation{School of Mathematical Sciences, Rochester Institute of Technology, Rochester, NY 14623, USA}

\date{\today}

\begin{abstract}
An analytic solution is obtained to the SEIR Epidemic Model.  The solution is created by constructing a single second-order nonlinear differential equation in $\ln S$ and  analytically continuing its divergent power series solution such that it matches the correct long-time exponential damping of the epidemic model.  This is achieved through an asymptotic approximant (Barlow et. al, 2017, Q. Jl Mech. Appl. Math, 70 (1), 21-48) in the form of a modified symmetric Pad\'e approximant that incorporates this damping. The utility of the analytical form is demonstrated through its application to the COVID-19 pandemic.

\end{abstract}

\maketitle




Asymptotic approximants have been successful at providing analytical solutions to many problems in mathematical
physics~\cite{BarlowJCP,BarlowAIChE,Barlow2015,Barlow:2017,Barlow:2017b,Beachley,Belden,BarlowWeinstein}.  Like the well-known Pad\'e approximant~\cite{BakerGammel,Bender}, they are
constructed to match a primary series expansion in a given region up to any specified order.  Unlike Pad\'e approximants, however, the form of an asymptotic approximant is not limited to a ratio
of polynomials, and its structure is chosen to enforce the asymptotic equivalence in a region away from the primary series expansion.  By increasing the number of terms in an
asymptotic approximant, it converges to the exact solution in these two regions \--- as well as at all points in between.  Convergence is certainly a necessary condition for a valid
approximant; although there is yet no proof, convergent approximants match the numerical solutions of systems examined
thus far~\cite{BarlowJCP,BarlowAIChE,Barlow2015,Barlow:2017,Barlow:2017b,Beachley,Belden,BarlowWeinstein}.

The COVID-19 outbreak motivates the application of asymptotic approximants to epidemiology models.  The method has seen recent success in providing a closed-form
solution to the Susceptible--Infected--Recovered (SIR) model~\cite{BarlowWeinstein}.  Here, we extend the method to the commonly  used Susceptible--Exposed--Infected--Recovered (SEIR)
model.  This model is formulated as a system of nonlinear ordinary differential equations, for which no exact analytic solution has yet been found.  The analytic nature of the
asymptotic approximant derived in what follows is advantageous, in that the accuracy and computational expense are not affected by the duration of the epidemic prediction; the form is
built such that it is accurate in $t\in[0,\infty)$ and can be evaluated at any specific time without the need for numerical marching. Depending on the duration, it may be beneficial to replace a numerical solution with the approximant within a
fitting algorithm that extracts SEIR parameters.   En route to the approximant, we also present an alternative formulation of the SEIR model as a single 2nd-order nonlinear
differential equation in $\ln S$.  This form enables an efficient series solution about $t=0$, asymptotic expansion as $t\to\infty$, and may itself prove attractive for future analysis.

The SEIR epidemic model considers the time-evolution of a susceptible population, $S(t)$, interacting with an exposed population, $E(t)$, and infected population,
$I(t)$, where $t$ is time.  This model is expressed as~\cite{Kermack}
\begin{subequations}
\begin{equation}
\frac{dS}{dt}=-\beta SI
\label{eq:S}
\end{equation}
\begin{equation}
\frac{dE}{dt}=\beta SI -\alpha E
\label{eq:E}
\end{equation}
\begin{equation}
\frac{dI}{dt}=\alpha E-\gamma I,
\label{eq:I}
\end{equation}
with a removed population (recovered + deaths), $R(t)$, evolved by
\begin{equation}
\frac{dR}{dt}=\gamma I
\label{eq:R}
\end{equation}
and constraints
\begin{equation}
S=S_0,~E=E_0,~I=I_0,~R=R_0\text{ at }t=0.
\label{eq:constraints}
\end{equation}
\label{eq:SEIR}
\end{subequations}
In~\neqref{eq:SEIR}, $\beta$, $\alpha$, $\gamma$, $S_0$, $E_0$, $I_0$, and $R_0$ are non-negative constant parameters~\cite{Kermack}.   Along with initial conditions
from~\neqref{eq:constraints}, the solution for $S$, $E$, and $I$ may be first obtained from~\neqref{eq:S} through~\neqref{eq:I} and the solution for
$R$ subsequently extracted using~\neqref{eq:R}.

We now manipulate the system~\neqref{eq:SEIR} into an equivalent 2{nd}-order equation in $\ln S$ to simplify the analysis that follows. Equations~\neqref{eq:S} and~\neqref{eq:E} are added to obtain
\begin{equation}
\frac{dS}{dt}+\frac{dE}{dt}=-\alpha E.
\label{eq:added}
\end{equation}
Solving~\neqref{eq:I} for $E$ and substituting into~\neqref{eq:added} then leads to
\begin{equation}
\frac{d^2I}{dt^2}+(\gamma+\alpha)\frac{dI}{dt}+\alpha\frac{dS}{dt}+\alpha\gamma I=0.
\label{eq:IODE}
\end{equation}
\eqref{eq:S} is rewritten as
\begin{equation}
I=-\frac{1}{\beta}\frac{d\ln S}{dt}
\label{eq:logform}
\end{equation}
and substituted into~\neqref{eq:IODE} to arrive at the 3{rd}-order equation
\begin{equation}
\frac{d^3\ln S}{dt^3}+(\gamma+\alpha)\frac{d^2\ln S}{dt^2}-\alpha\beta\frac{dS}{dt}+\alpha\gamma\frac{d\ln S}{dt}=0.
\label{eq:thirdorder}
\end{equation}
Equation~\eqref{eq:thirdorder} may be integrated to yield
\begin{equation}
\frac{d^2\ln S}{dt^2}+(\gamma+\alpha)\frac{d\ln S}{dt}-\alpha\beta S+\alpha\gamma \ln S=C,
\label{eq:secondorder}
\end{equation}
where the integration constant
\begin{subequations}
\begin{equation}
C=\alpha\gamma\ln(S_0)-\alpha\beta\left(E_0+I_0+S_0\right)
\label{eq:C}
\end{equation}
is obtained by evaluating the left-hand side of~\neqref{eq:secondorder} at $t=0$ using~\neqref{eq:I},~\neqref{eq:constraints}. and~\neqref{eq:logform}. The form of~\eqref{eq:secondorder} suggests that  the variable substitution $f=\ln S$ be made, and the result is
\begin{equation}
\frac{d^2f}{dt^2}+(\gamma+\alpha)\frac{df}{dt}-\alpha\beta e^f +\alpha\gamma f=C
\label{eq:singleODE}
\end{equation}
where, from~\neqref{eq:constraints} and~\neqref{eq:logform},
\begin{equation}
f=\ln S_0,~\frac{df}{dt}=-\beta I_0\text{ at }t=0.
\label{eq:ICs}
\end{equation}
\label{eq:newODE}
\end{subequations}
Once~\neqref{eq:newODE} is solved for $f$, $S$ is extracted as:
\begin{subequations}
\begin{equation}
S=e^f.
\label{eq:Seqn}
\end{equation}
The solution for $I$ follows directly from~\neqref{eq:logform} and~\neqref{eq:Seqn} as
\begin{equation}
I=-\frac{1}{\beta}\frac{df}{dt}.
\label{eq:Ieqn}
\end{equation}
After substituting~\neqref{eq:Ieqn} into~\neqref{eq:R}, integrating, and applying the constraint~\neqref{eq:ICs}, $R$ is expressed as:
\begin{equation}
R=R_0-\frac{\gamma}{\beta}\left(f-\ln S_0\right).
\label{eq:Reqn}
\end{equation}
Lastly, the conservation of $S+E+I+R$ provides a solution for $E$ as
\begin{equation}
E=E_0+I_0+S_0+R_0-I-S-R,
\label{eq:Eeqn}
\end{equation}
as seen by adding equations~\neqref{eq:S} through~\neqref{eq:R}, integrating in $t$, and applying~\neqref{eq:constraints}.
\label{eq:SEIReqns}
\end{subequations}
\par

 The series solution of~\neqref{eq:newODE} is given by
\begin{subequations}
\begin{equation}
f=\sum_{n=0}^\infty a_n t^n,~a_0=\ln S_0,~a_1=-\beta I_0
\label{eq:series}
\end{equation}
\begin{equation}
a_2=\left[C-\left(\alpha+\gamma\right)a_1+\alpha\beta S_0-\alpha\gamma a_0\right]/2
\label{eq:a2}
\end{equation}
\begin{equation}
a_{n+2}=\frac{\alpha\beta \tilde{a}_n-(\gamma+\alpha)(n+1)a_{n+1}-\alpha\gamma a_n}{(n+2)(n+1)},~n>0
\label{eq:coefficients}
\end{equation}
\begin{equation}
\tilde{a}_{n>0}=\frac{1}{n+1}\sum_{j=0}^{n} (n-j+1)a_{n-j+1}\tilde{a}_{j},~~\tilde{a}_0=S_0.
\label{eq:at}
\end{equation}
\label{eq:SeriesSolution}
\end{subequations}
The result~\neqref{eq:SeriesSolution} is obtained by the standard procedure of inserting~\neqref{eq:series} into~\neqref{eq:newODE} and finding a recursion for the coefficients by equating like-terms.  It is thus necessary to obtain the expansion of the nonlinear term $e^f\equiv S$ in~\neqref{eq:newODE}.  To do so, we solve for the coefficients of $S=\sum \tilde{a}_n t^n$ by applying Cauchy's product rule to the chain-rule result $f'S=S'$ and evaluating like-terms; this leads to the recursive expression given by~\neqref{eq:at}.   Although the series solution given by~\neqref{eq:SeriesSolution} is an analytic solution to~\neqref{eq:newODE}, it is only valid within its radius of convergence and is incapable of capturing the long-time behavior of the system.  This motivates the use of an approximant to analytically continue the series beyond this radius.

The long-time asymptotic behavior of the system~\neqref{eq:newODE} is required to develop our asymptotic approximant, and so we proceed as follows.  It has been proven in prior
literature~\cite{hethcotechapter} that $S$ approaches a limiting value, $S_\infty$, as $t\to\infty$, and this corresponds to $I\to0$ in the same limit.   Thus,
$f$ approaches a limiting value, $f_\infty\equiv \ln S_\infty$, as $t\to\infty$.  The value of $f_\infty$ satisfies the following equation~\cite{hethcotechapter} 
\begin{subequations}
\begin{equation}
e^{f_\infty}-\frac{\gamma}{\beta}\left(f_\infty-\ln S_0\right)-E_0-I_0-S_0 =0
\label{eq:infinite}
\end{equation} 
in the interval
\begin{equation}
f_\infty\in(-\infty,\ln \gamma/\beta).
\label{eq:bounds}
\end{equation}
\end{subequations}

We expand $f$ as $t\to \infty $ as follows:
\begin{equation}
 f\sim f_{\infty } +g(t)\text{ where }g\to 0\text{ as }t\to \infty. 
\label{eq:perturb}
\end{equation}
\eqref{eq:perturb} is substituted into~\neqref{eq:singleODE} (with~\neqref{eq:C}), $e^{g}$ is replaced with its power series expansion, and terms of $O(g^{2})$ are neglected to achieve the following linearized equation
\begin{equation}
\frac{d^2 g}{dt^2}+(\gamma+\alpha)\frac{dg}{dt}+\left(\alpha\gamma-\alpha\beta e^{f_\infty}\right)g=0.
\label{eq:linearODE}
\end{equation}
The general solution to~\neqref{eq:linearODE} is
\begin{subequations}
\begin{equation}
 g =\epsilon_1 e^{\lambda_1 t} +\epsilon_2 e^{\lambda_2 t}
  \label{eq:f1}
\end{equation}
\begin{equation}
 \lambda_{1,2}=\frac{1}{2}\left[-\alpha-\gamma\pm\sqrt{(\gamma-\alpha)^2+4\alpha\beta e^{f_\infty}}\right]
   \label{eq:lambda}
\end{equation}
 \label{eq:linear}
\end{subequations}
where $\epsilon_1$ and $\epsilon_2$ are unknown constants and $\lambda_2<\lambda_1<0$ since $e^{f_\infty}<\gamma/\beta$ from~\neqref{eq:bounds}.  Thus the long-time asymptotic behavior of $f$ is given by
\begin{equation}
f\sim f_\infty+\epsilon_1 e^{\lambda_1 t},~t\to\infty.
\label{eq:asymptotic}
\end{equation}

 Higher order corrections to the expansion~\neqref{eq:asymptotic} may be obtained by the method of dominant balance~\cite{Bender} as a series of more rapidly damped exponentials.
However, the pattern by which the corrections are asymptotically ordered is not as straightforward as that of the SIR model, provided in~\citet{BarlowWeinstein}.  In that work, an
asymptotic approximant is constructed as a series of exponentials that exactly mimics the long-time expansion.  In the SEIR model, complications in the higher-order asymptotic
behavior arise from the competition between the two exponentials in~\neqref{eq:f1}. Here, we enforce the leading-order $t\to\infty$  behavior given by~\neqref{eq:asymptotic} and make
a more traditional choice for matching with the $t=0$ expansion~\neqref{eq:SeriesSolution}.  We create an approximant with an embedded rational function with equal-order
numerator and denominator (i.e.,~a symmetric Pad\'e approximant~\cite{Bender}), such that it approaches the unknown constant $\epsilon_1$ in~\neqref{eq:asymptotic} as $t\to\infty$,
while converging to the intermediate behavior at shorter times. The assumed SEIR approximant is given by
\begin{equation}
 f_{A,N}=f_\infty+e^{\lambda_1 t}\frac{\displaystyle\sum_{n=0}^{N/2} A_n t^n}{1+\displaystyle\sum_{n=1}^{N/2} B_n t^n},
 \text{$N$ even}
\label{eq:Aform}
\end{equation}
where the $A_n$ and $B_n$ coefficients are obtained such that the Taylor expansion of~\neqref{eq:Aform} about $t=0$ is exactly~\neqref{eq:SeriesSolution}.    Note that,
although a rational function is being used in~\neqref{eq:Aform}, it is not a Pad\'e approximant itself.  Pad\'e approximants are only capable of capturing $t^n$ behavior in the long-time
limit, where $n$ is an integer. The pre-factor $e^{\lambda_1 t}$ is required to make~\neqref{eq:Aform} an \textit{asymptotic} approximant for the SEIR model.   However, we may
still make use of fast Pad\'e coefficient solvers~\cite{padeapprox,Trefethen} by recasting~\neqref{eq:Aform} as a Pad\'e approximant for the series that results from the Cauchy product between
the expansions of $e^{-\lambda_1 t}$ and $f-f_\infty$, expressed as
\begin{equation}
\sum_{n=0}^N\left[\sum_{j=0}^n\frac{\left(-\lambda_1\right)^j}{j!} \tilde{a}_{n-j} \right]t^n=\frac{\displaystyle\sum_{n=0}^{N/2} A_n t^n}{1+\displaystyle\sum_{n=1}^{N/2} B_n t^n},
\label{eq:Pade}
\end{equation}
where $\tilde{a}_0=a_0-f_\infty$ and $\tilde{a}_{n>0}=a_{n>0}$.   A MATLAB
code to compute the $A_n$ and $B_n$ coefficients of~\neqref{eq:Aform} (for given $\alpha$, $\gamma$, $\beta$, $S_0$, $E_0$, $I_0$) is
available from the authors~\cite{SEIRcode}.

\begin{figure*}
\begin{tabular}{c}
\includegraphics[width=3.5in]{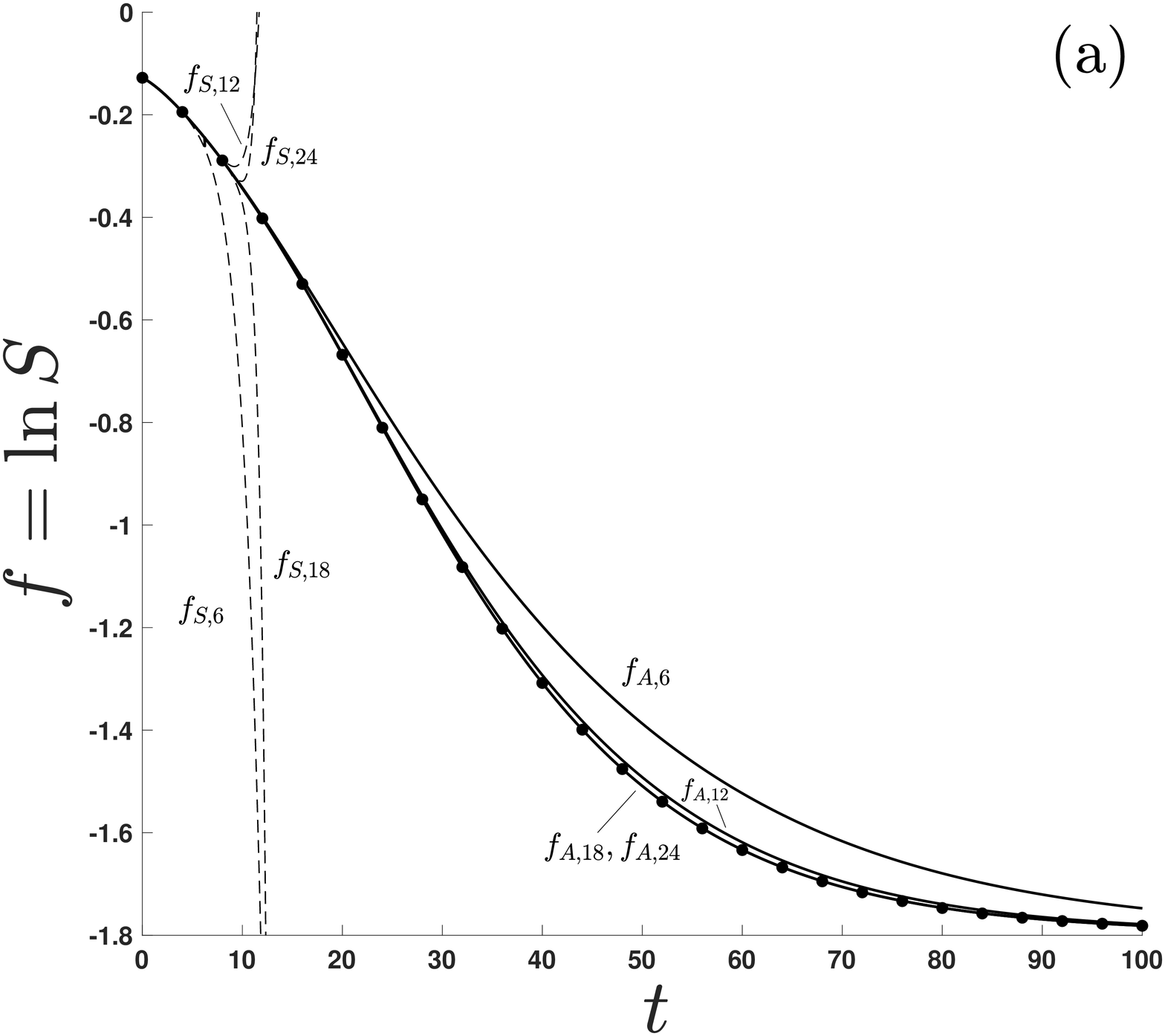} 
\includegraphics[width=3.5in]{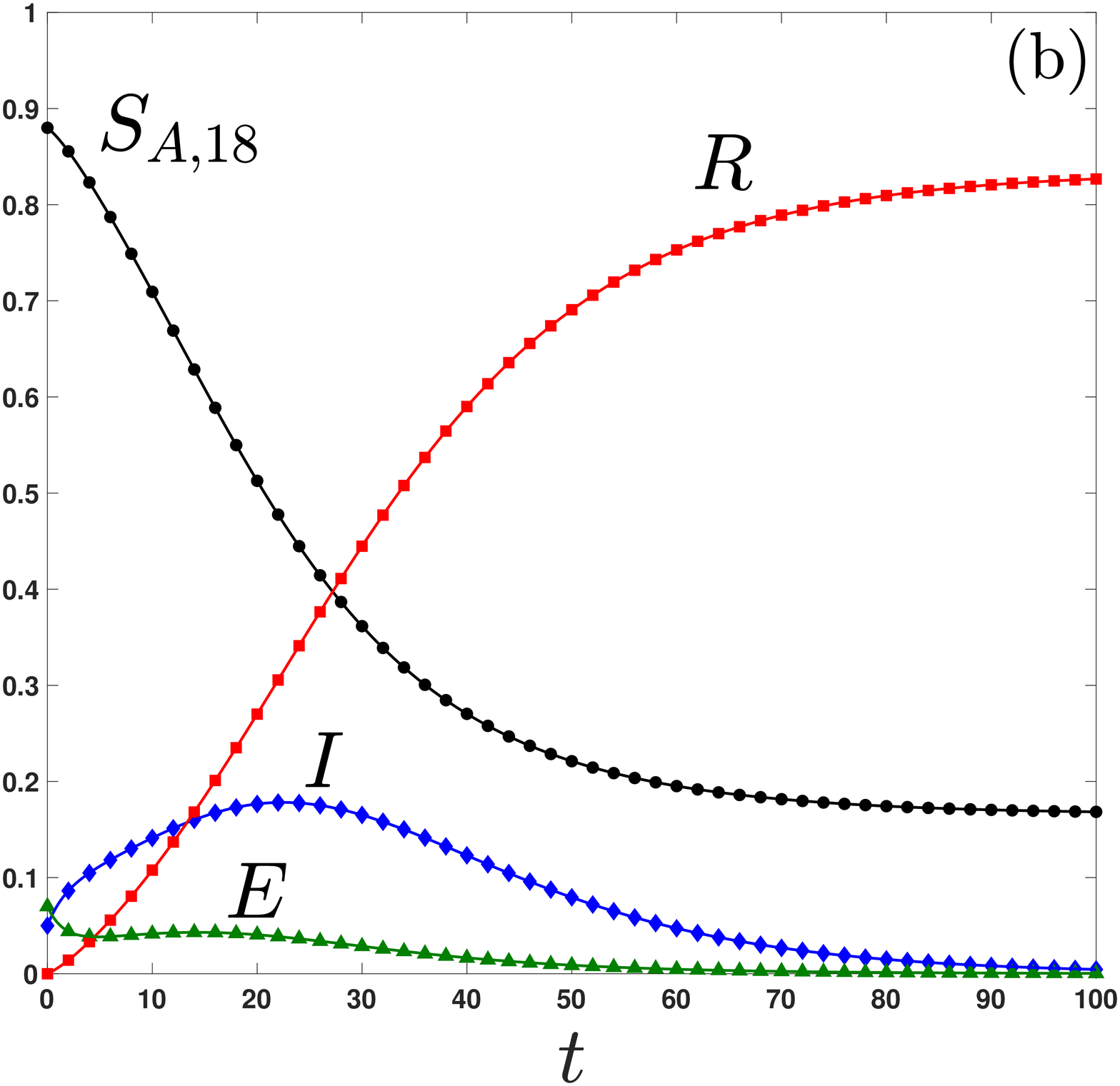} 
\end{tabular}
\caption{Analytical and numerical solutions to the SEIR model~\neqref{eq:SEIR}, where the susceptible ($S$), exposed ($E$), infected ($I$), and recovered
($R$) populations are represented as a fraction of the total population and $t$ is in units of days.  (a)  Solution shown in terms of $f\equiv \ln S$.  As the number of
terms $N$ is increased, the series solution, denoted by $f_{S,N}$ (given by~\neqref{eq:SeriesSolution}, dashed curves), diverges and the approximant, denoted by $f_{A,N}$
(given by~\neqref{eq:Aform}, solid curves), converges to the exact (numerical) solution ($\bullet$'s). Corresponding relative errors are provided in figure~\ref{fig:error}a. (b)
The converged asymptotic approximant for $f$ is used to obtain $S$, $E$, $I$, and $R$ from~\neqref{eq:SEIReqns} shown by solid curves and
compared with the numerical solution (closed symbols).  The model parameters values and initial conditions $\alpha=0.466089$, $\beta=0.2$, $\gamma=0.1$, $S_0=0.88$, $E_0=0.07$,
$I_0=0.05$,  and $R_0=0$ are taken from estimates of Ebola virus propagation examined in~\citet{Rachah}.}
\label{fig:Ebola}
\end{figure*}

\begin{figure*}
\begin{tabular}{c}
\includegraphics[width=3.5in]{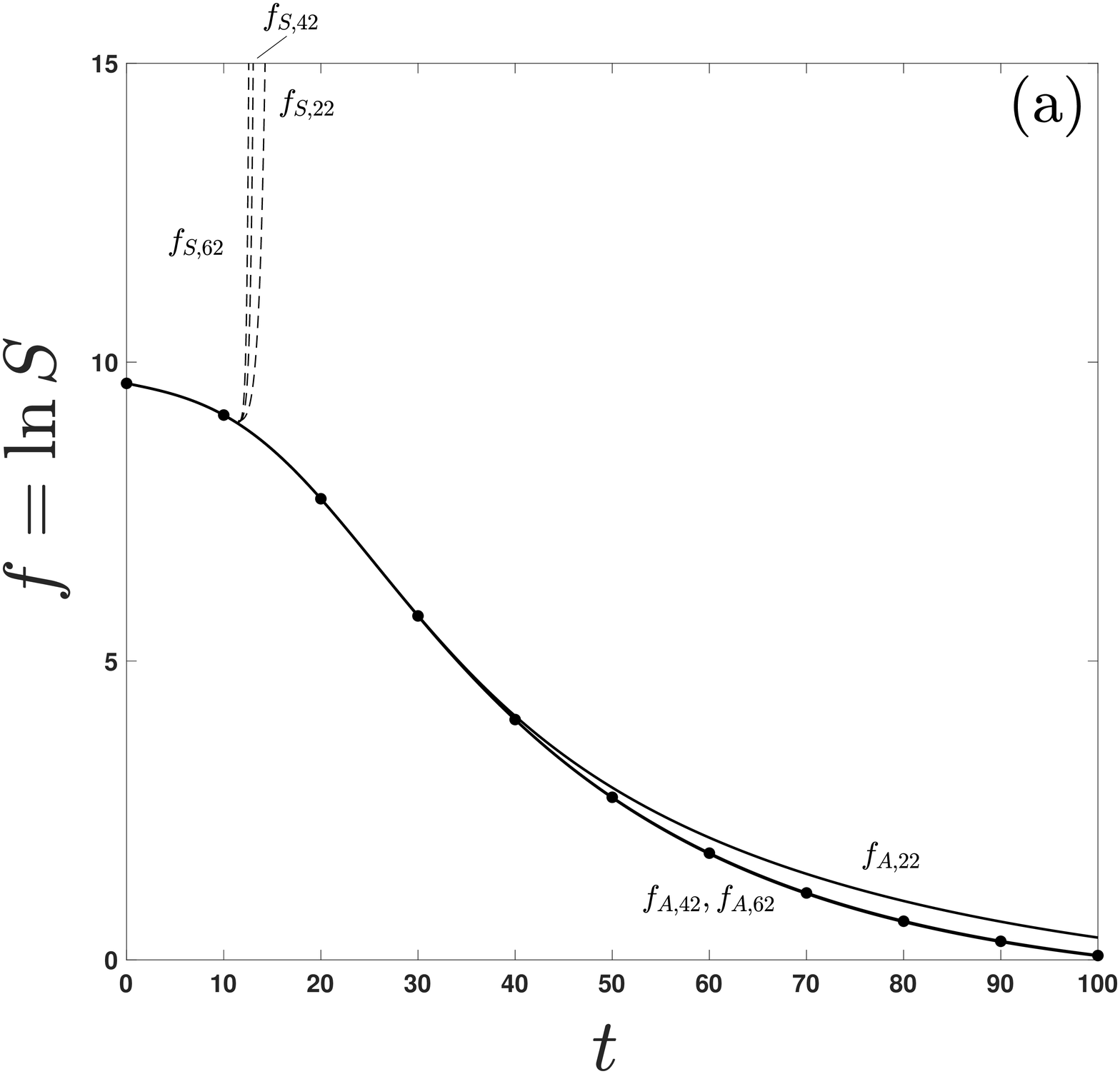} 
\includegraphics[width=3.5in]{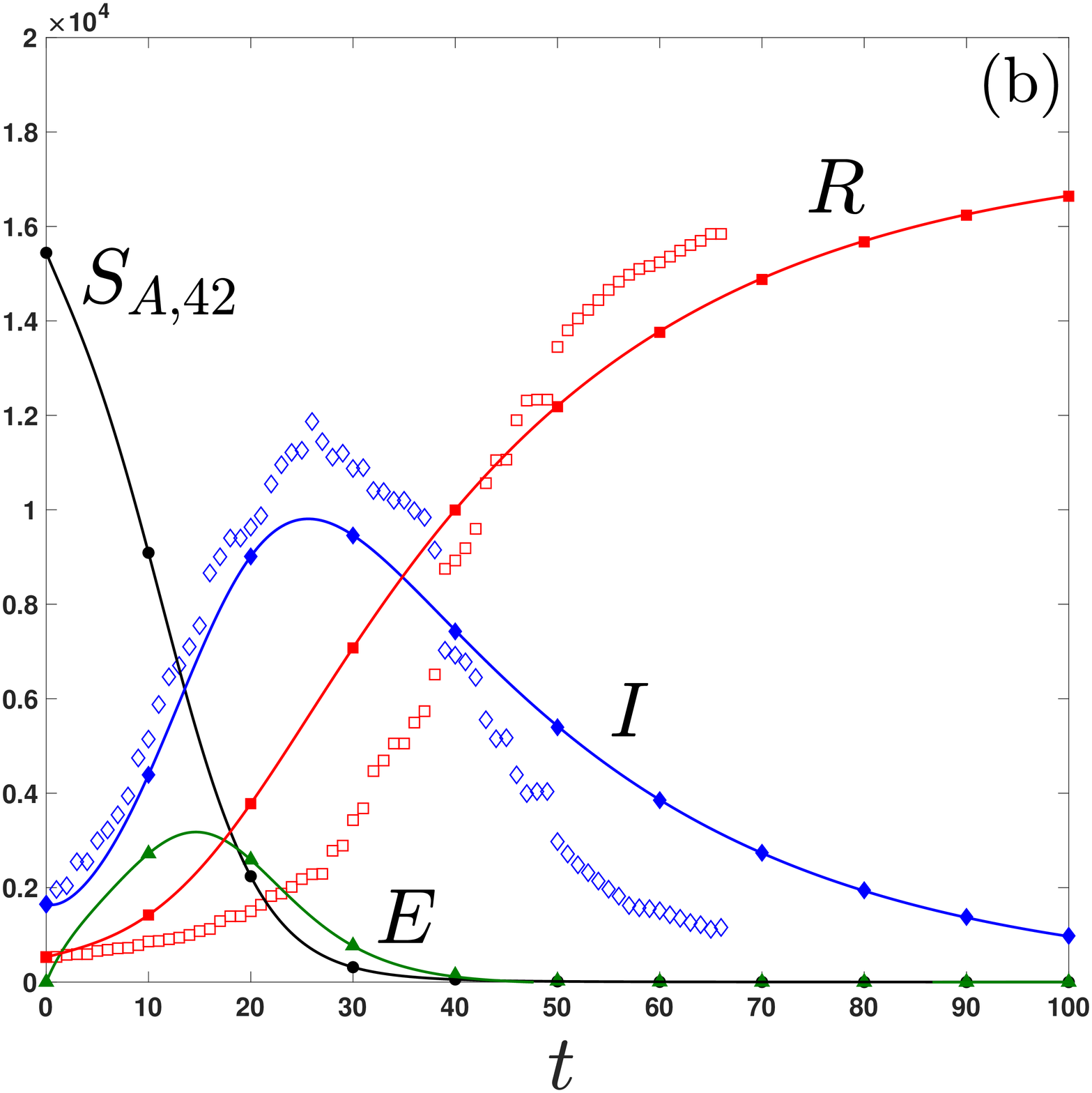} 
\end{tabular}
\caption{Analytical and numerical solutions to the SEIR model~\neqref{eq:SEIR}, where $S$, $E$, $I$, $R$ are in units of people and $t$
is in days.  All other notation and labels are the same as in figure~\ref{fig:Ebola}, except $R$ now also includes deaths.  Corresponding relative errors are provided in figure~\ref{fig:error}c. SEIR model parameters values and unknown initial conditions are obtained via a least-squares fit to the Japan COVID-19 outbreak data~\cite{covid} (open
symbols).  Best fit parameters are $\alpha$=0.2332207, $\beta$=2.040015$\times10^{-5}$, $\gamma$=0.034334, $S_0$=15442, and $E_0$=0.  The initial conditions
$I_0=1649$ and $R_0$=529  are taken directly from the data set~\cite{covid} at a chosen $t=0$ (here April 1, 2020).}
\label{fig:Japan}
\end{figure*}

\begin{figure*}
\begin{tabular}{cc}
\includegraphics[width=3.5in]{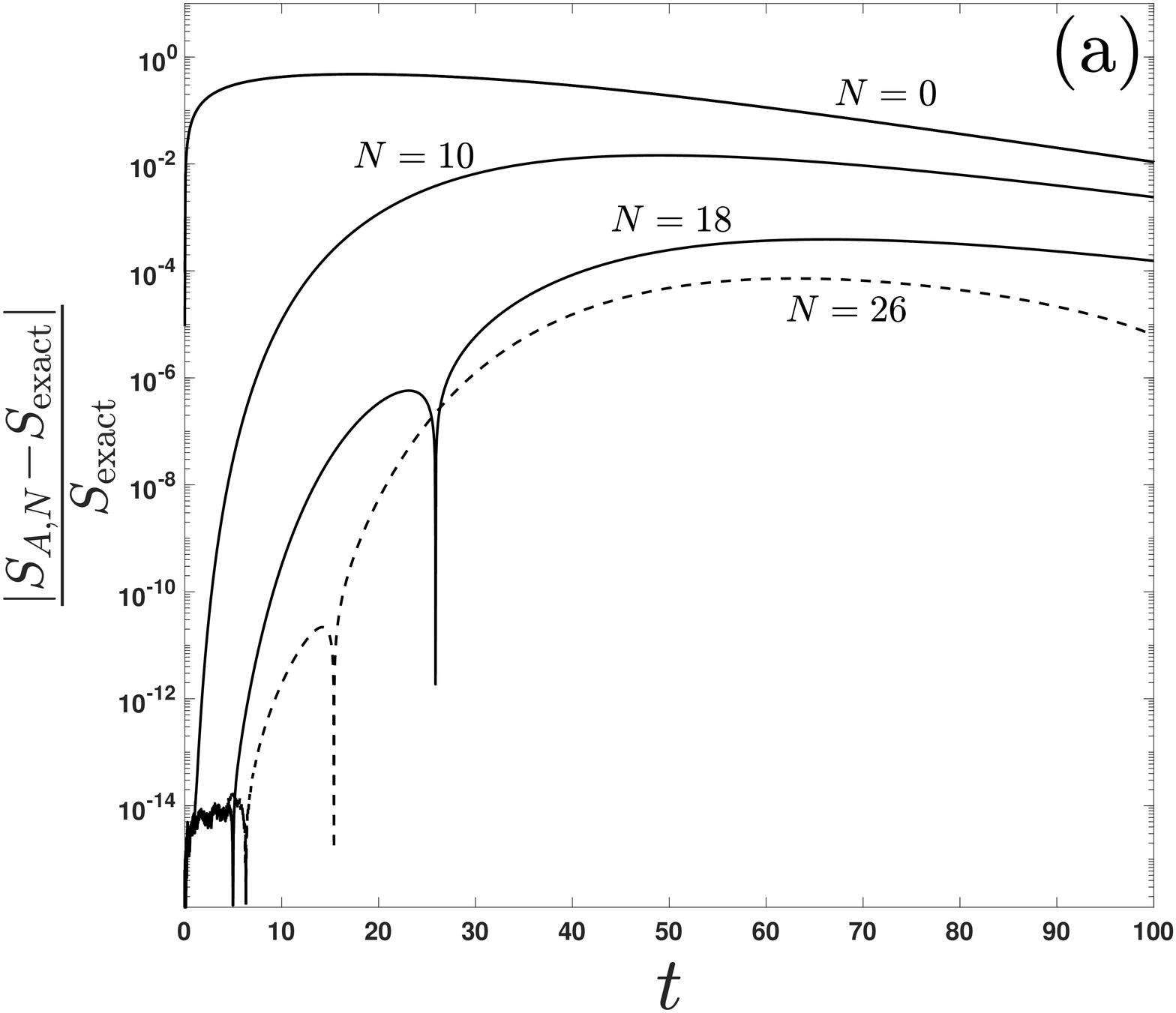} & \includegraphics[width=3.5in]{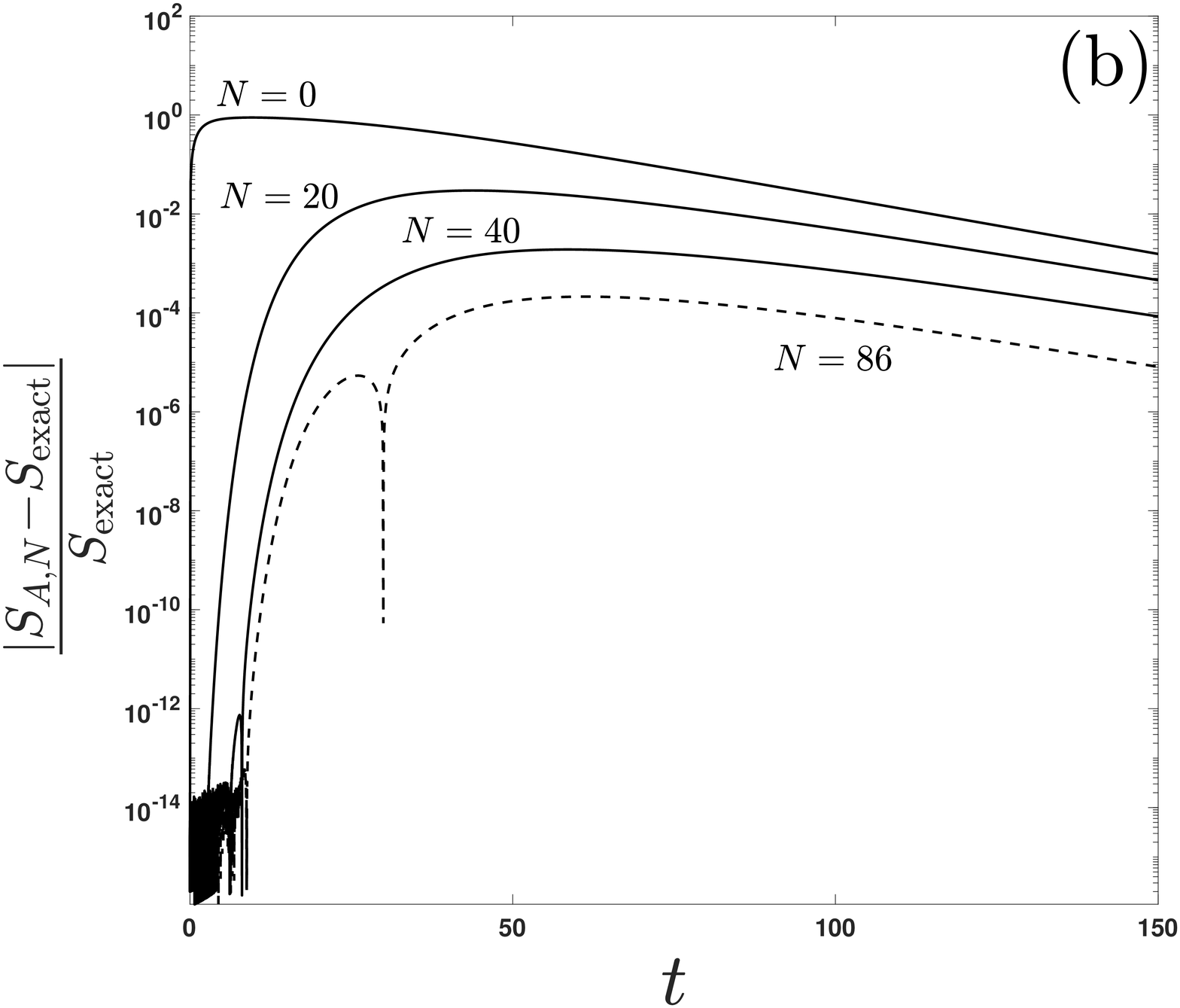}\\
\includegraphics[width=3.5in]{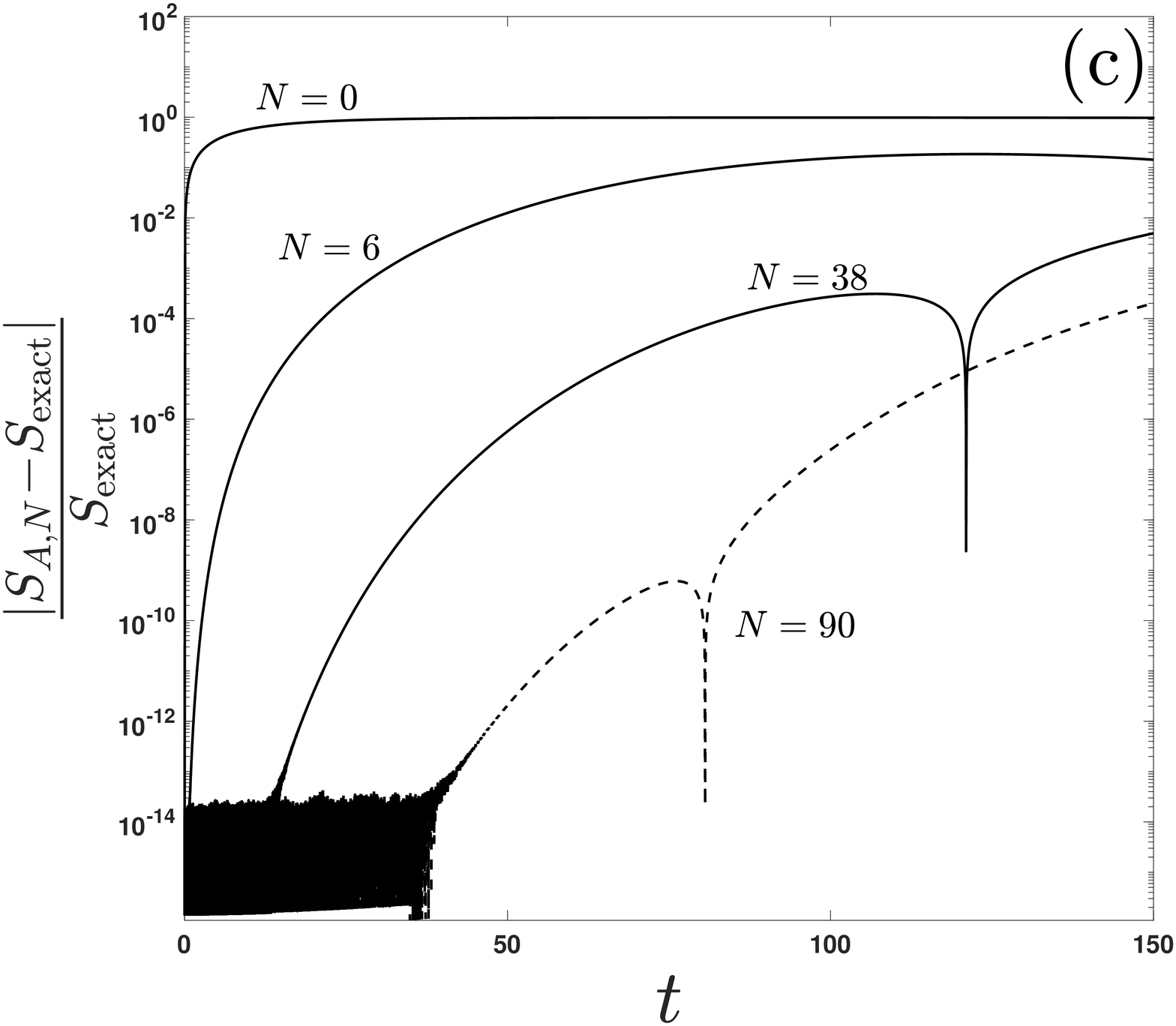} & \includegraphics[width=3.5in]{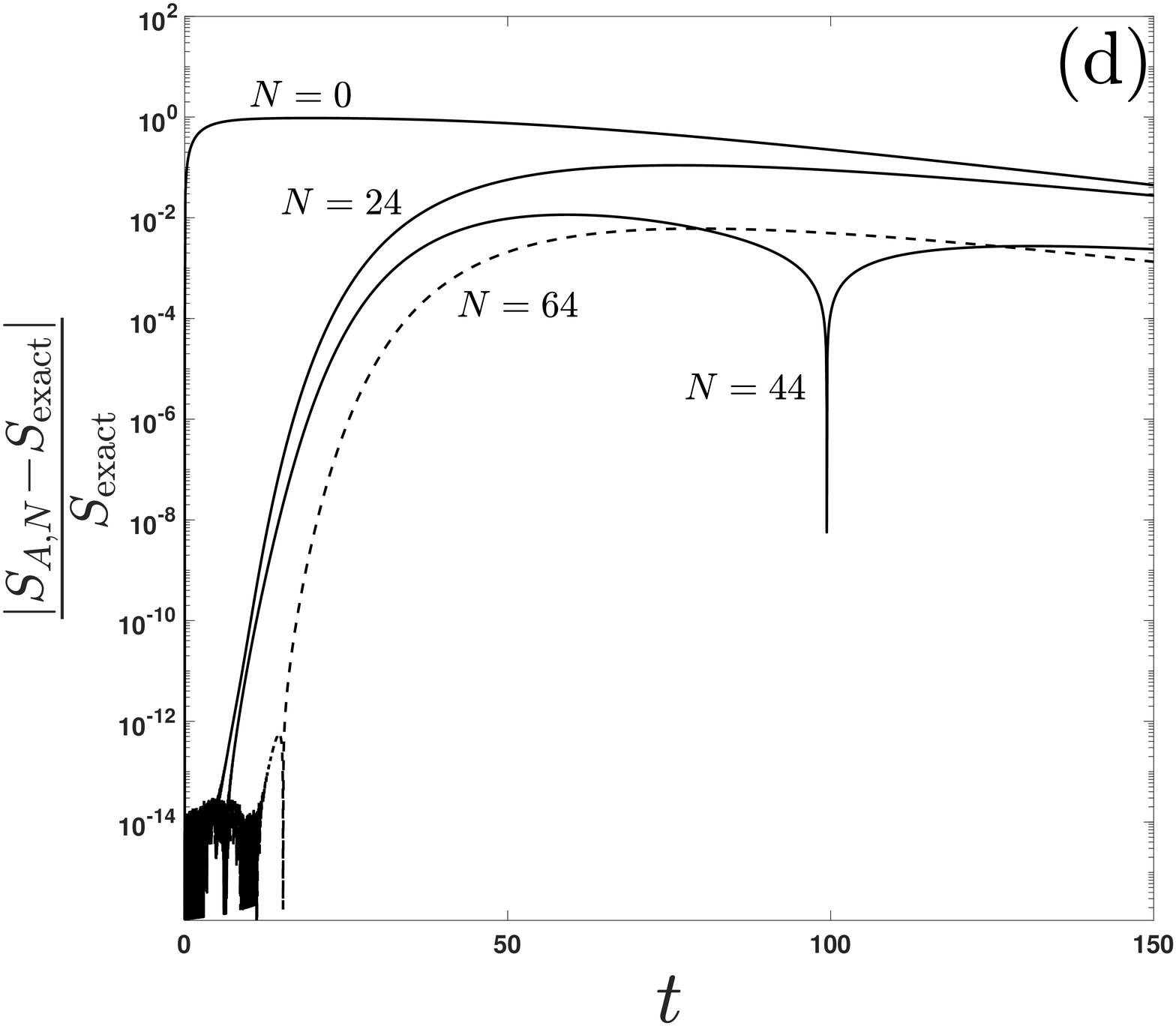}
\end{tabular}
\caption{Relative error of the approximant~\neqref{eq:Aform} for increasing $N$ as a function of  $t$ (in days).  The exact solution is taken to be the numerical solution of~\neqref{eq:SEIR}, computed using the 4{th}-order Runge--Kutta scheme with a time-step of 10$^{-4}$.  The subfigures (a)\--(d) correspond to the cases presented in figure~\ref{fig:Ebola}\--\ref{fig:Japan}, respectively.  For all figures, $N$ is taken up until optimal truncation is achieved, indicated by a dashed curve.  The cusps in the figures have no physical meaning and simply indicate where the sign of $(S_{A,N}-S_{\mathrm{exact}})$ changes.}
\label{fig:error}
\end{figure*}

The SEIR approximant~\neqref{eq:Aform} is thus an analytic expression that, by construction, matches the correct $t\to\infty$ behavior given by~\neqref{eq:asymptotic} and whose expansion about $t=0$ is exact to $N${th}-order.  A comparison between the approximant solution~\neqref{eq:Aform} and the numerical solution to~\neqref{eq:SEIR} is provided in figures~\ref{fig:Ebola}\--\ref{fig:Japan} with the relative error for all four cases provided in figure~\ref{fig:error}. The indicated error in figure~\ref{fig:error} is calculated by comparing $S(t)$ to its accurate numerical solution (assumed to be exact); curves showing the same order of accuracy are obtained when the other dependent variables of the model are examined.

\begin{figure*}
\begin{tabular}{c}
\includegraphics[width=3.5in]{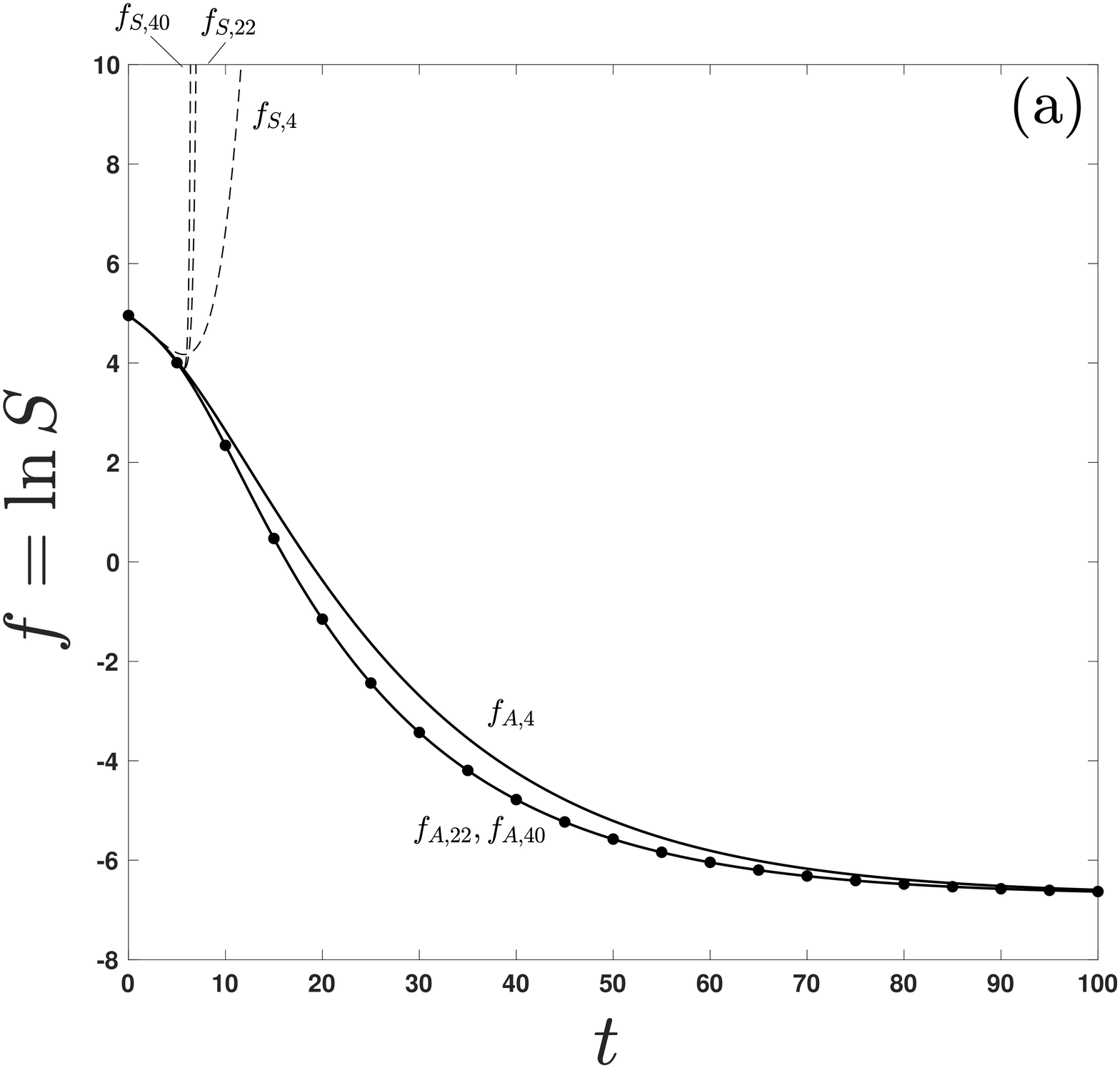} 
\includegraphics[width=3.5in]{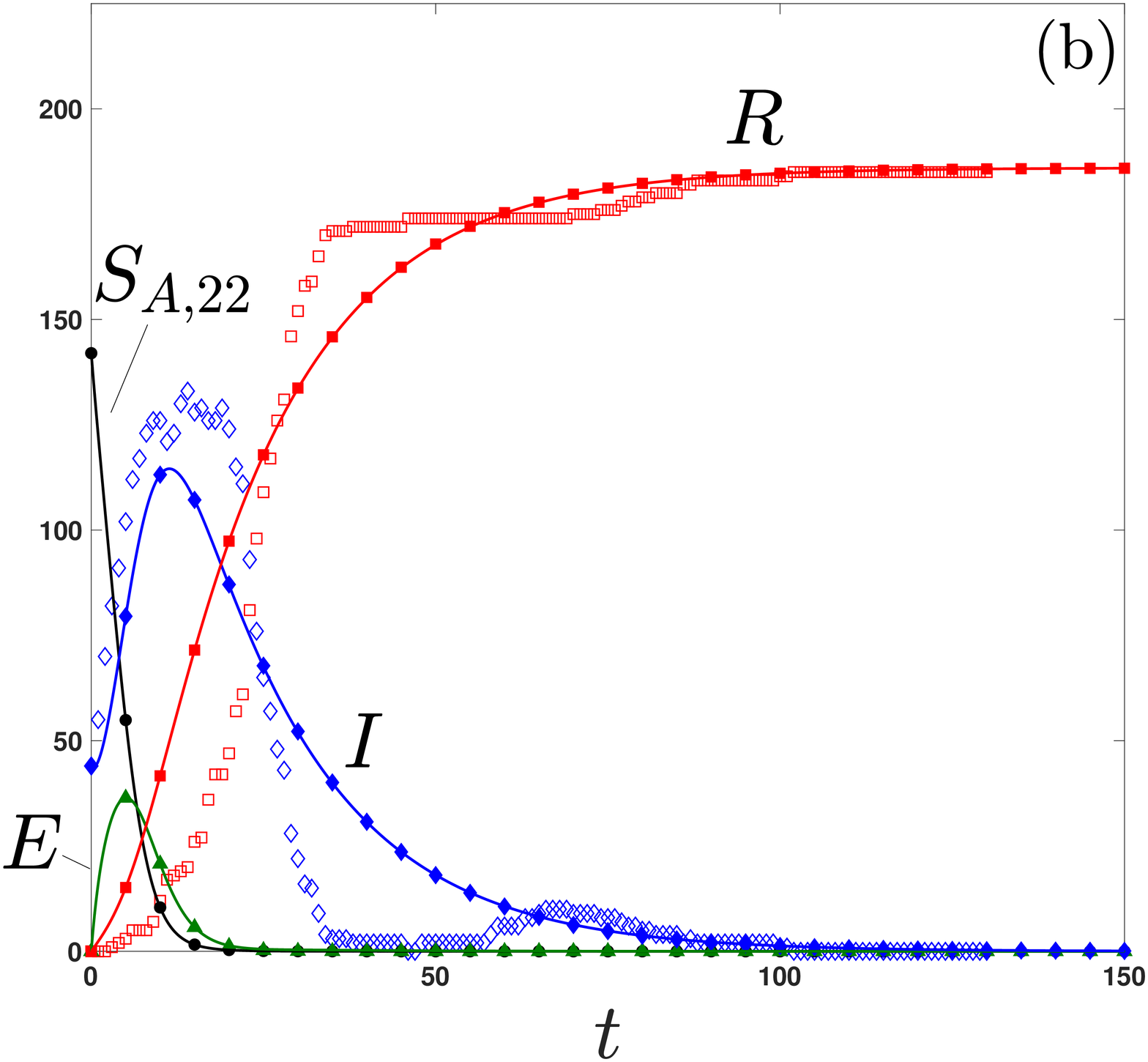} 
\end{tabular}
\caption{Analytical and numerical solutions to the SEIR model~\neqref{eq:SEIR}, where $S$, $E$, $I$, $R$ are in units of people and $t$
is in days.  All other notation and labels are the same as in figure~\ref{fig:Ebola}, except $R$ now also includes deaths. Corresponding relative errors are provided in figure~\ref{fig:error}b.  SEIR model parameters values and unknown initial conditions are obtained via a least-squares fit to the Yunan, China COVID-19 outbreak data~\cite{covid}
(open symbols).  Best fit parameters are $\alpha$=0.395031, $\beta$=0.00333, $\gamma$=0.0553093, $S_0$=142, and $E_0$=0.  The initial conditions
$I_0=44$ and $R_0$=0 are taken directly from the data set~\cite{covid} at a chosen $t=0$ (here January 28, 2020).}
\label{fig:Yunan}
\end{figure*}

\begin{figure*}
\begin{tabular}{c}
\includegraphics[width=3.5in]{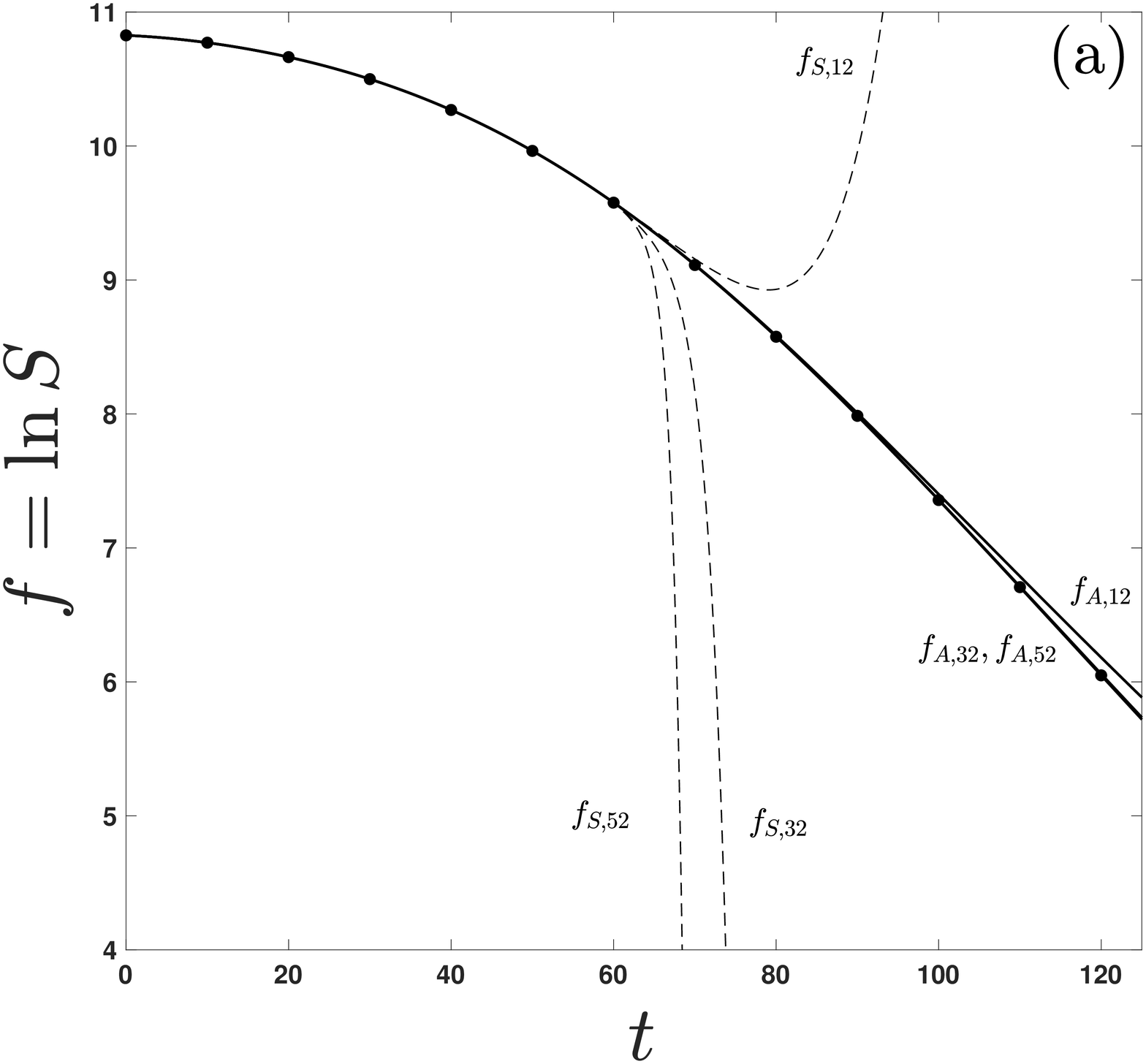} 
\includegraphics[width=3.5in]{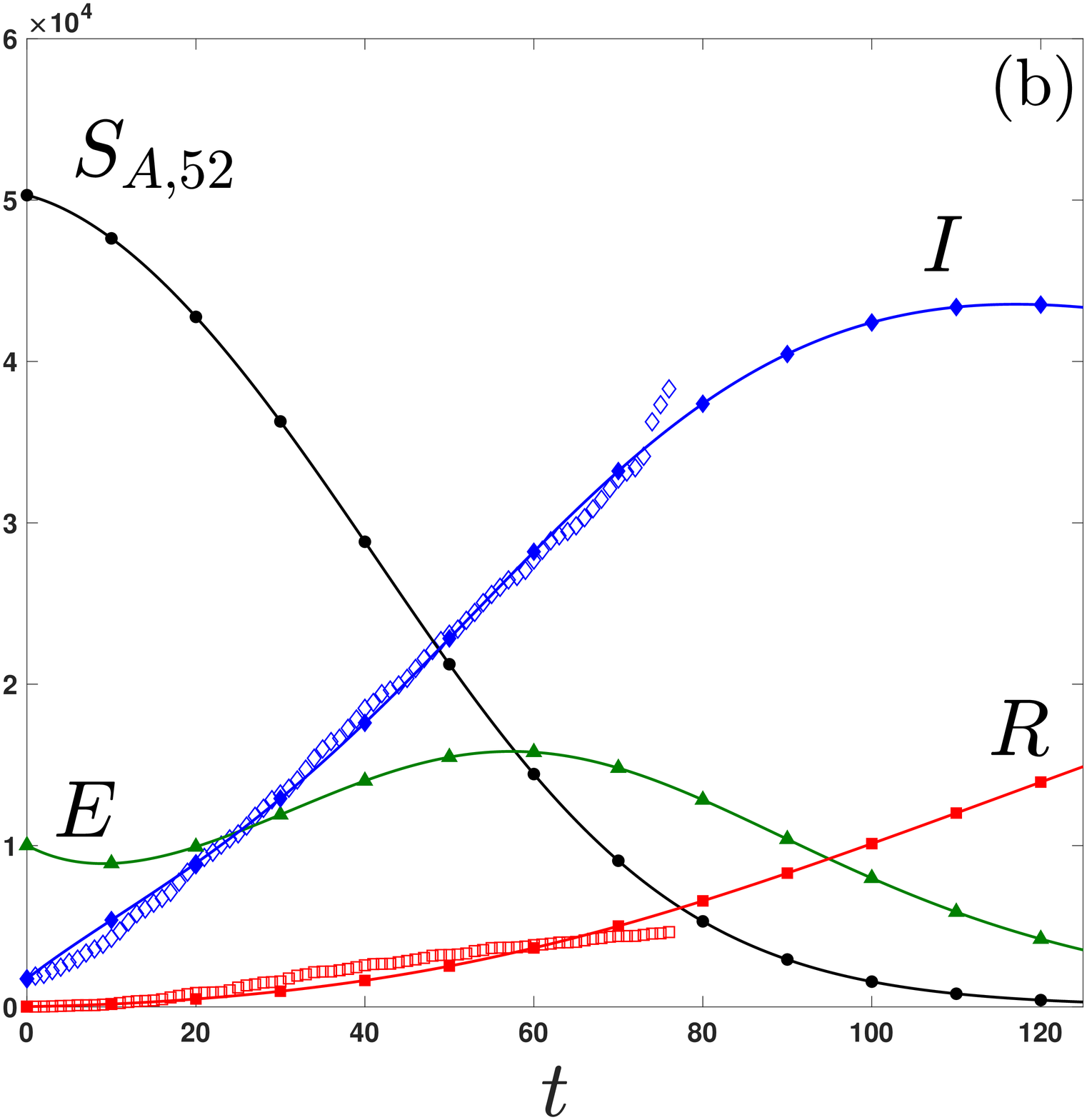} 
\end{tabular}
\caption{Analytical and numerical solutions to the SEIR model~\neqref{eq:SEIR}, where $S$, $E$, $I$, $R$ are in units of people and $t$
is in days.  All other notation and labels are the same as in figure~\ref{fig:Ebola}, except $R$ now also includes deaths. Corresponding relative errors are provided in figure~\ref{fig:error}c. SEIR model parameters values and unknown initial conditions are obtained via a least-squares fit to the Sweden COVID-19 outbreak data~\cite{covid} (open
symbols).  Best fit parameters are $\alpha$=0.041281, $\beta$=1.513332$\times10^{-6}$, $\gamma$=0.004407, $S_0$=50306, and $E_0$=10015.  The initial
conditions $I_0=1743$ and $R_0$=20  are taken directly from the data set~\cite{covid} at a chosen $t=0$ (here March 21, 2020).}
\label{fig:Sweden}
\end{figure*}

Figure~\ref{fig:Ebola}a provides a typical comparison of the $N$-term series solution~\neqref{eq:SeriesSolution} denoted by $f_{S,N}$ (dashed lines), the $N$-term
approximant~\neqref{eq:Aform} denoted by $f_{A,N}$ (solid lines), and the numerical solution ($\bullet$'s).  Note that the series solution has a finite radius of convergence as
evidenced by the poor agreement and divergence from the numerical solution at larger times, even as additional terms are included.   By contrast, the approximant converges as
additional terms are included.  For $N=18$, the approximant is visibly indistinguishable from the numerical solution on the scale of figure~\ref{fig:Ebola}a.  Figure~\ref{fig:error}a
provides the relative error of the approximant for the data shown in figure~\ref{fig:Ebola}.  Increasing the number of terms beyond $N=18$ does improve accuracy up to a point,
but a minimum error barrier is eventually reached of $O$(10$^{-6}$) at $N=26$; note that, to make this assessment, we take the maximum relative error with
respect to time for each $N$ (the maxima in figure~\ref{fig:error}a).  For larger values of $N$, the maximum error increases, and the approximant begins to diverge, i.e.~there is an
optimal value of $N$ at which to truncate the approximant. Asymptotic approximants can exhibit an optimal truncation~\cite{Beachley,Belden} as is often observed
with asymptotic expansions in general~\cite{Bender}.  We emphasize here that a numerical solution is not needed to assess convergence of approximants to within their optimal
truncation; convergence in the Cauchy sense (i.e.,~the distance between approximants decreases with increasing $N$) may be examined.  In addition to this issue,  deficient
approximants are possible with increasing $N$ due to zeros that can arise in the denominator of~\neqref{eq:Aform}.  Such approximants are ignored in assessing convergence.
To avoid this behavior, the lowest number of terms that yields the desired accuracy should be chosen. The convergence of the approximant with increasing $N$ (up until its
optimal truncation) is a necessary condition for a valid approximant.   In figure~\ref{fig:Ebola}b, the converged ($N=18$) asymptotic approximant for $f$ is used to
obtain analytic solutions for $S$, $E$, $I$, and $R$ from~\neqref{eq:SEIReqns}, which are compared with the numerical solution for these
quantities.  The approximant for $N=18$ agrees with numerics within the visible scale of the plot, with errors quantified by figure~\ref{fig:error}a.

Figure~\ref{fig:Ebola} results described above correspond to a case examined in~\citet{Rachah} to model an Ebola outbreak. In figures~\ref{fig:Yunan},~\ref{fig:Sweden},
and~\ref{fig:Japan}, the approximant is applied to COVID-19 data~\cite{covid} for Yunan (China), Sweden, and Japan, respectively.  Figure \ref{fig:error}b\--d provide the relative
error for these cases; the largest indicated value of $N$ in each figure (corresponding to dashed curves) is the optimal truncation as discussed above for figure~\ref{fig:error}a.  Note that we extensively
surveyed the available COVID-19 data~\cite{covid}, and the results in figures~\ref{fig:Yunan}\--\ref{fig:error} are representative of the fits and variability in the number of terms needed for
convergence of the approximant up to its optimal truncation.  

Note that the reported COVID-19 outbreak data~\cite{covid} is provided in terms of \textit{confirmed} cases, \textit{recovered} individuals, and \textit{deaths} per day.  We use
\textit{recovered} + \textit{deaths} as an approximation to the removed population $R$ and use \textit{confirmed} $-$ \textit{recovered} $-$
\textit{deaths} as an approximation to $I$ in the SEIR model.    It is acknowledged that the actual COVID-19 data is influenced by effects not
included in the SEIR model, and this can affect the ability of the model to closely fit actual COVID-19 data. The data approximations made here are to enable comparisons with model
predictions.  The ability of the approximant to match numerical results is unaffected by such approximations.  Disagreement between the model and epidemic data after fitting is
attributed to the applicability of the SEIR model and not the approximant.

In figures~\ref{fig:Yunan}\--\ref{fig:Japan}, a  least squares fit to $I$ and $R$ data is used to extract SEIR parameters $\alpha$, $\beta$, $\gamma$
and initial conditions $S_0$ and $E_0$. To do so, the initial values of $I_0$ and $R_0$ are taken directly from the COVID-19 data set~\cite{covid}.
Additionally, the time $t=0$ is chosen such that disease has progressed to a point where initial trends are observed, so that curve shapes are consistent with those reasonably
predicted by the SEIR model.   Adjustments such as this have been well described in fits done in previous work~\cite{Peirlinck,Linka}.  The initial guesses for the iterative
least-squares fit are taken from data fits for earlier times than examined here~\cite{Peirlinck,Linka}.

Our results demonstrate that an asymptotic approximant can be used to provide accurate analytic solutions to the SEIR model.  Future work should examine the ability of the asymptotic
approximant technique to yield closed-form solutions for even more sophisticated epidemic models, as well as their endemic counterparts~\cite{hethcotechapter}.

%

\end{document}